\documentclass[screen,acmsmall]{acmart}
\usepackage{subcaption}
\usepackage{caption}
\usepackage{dirtytalk}
\usepackage{graphicx}
\usepackage{colortbl}
\usepackage{booktabs}
\usepackage{multirow}
\usepackage{balance}
\AtBeginDocument{%
  }

\setcopyright{cc}
\setcctype{by-nc-nd}
\acmJournal{PACMHCI}
\acmYear{2026} \acmVolume{10} \acmNumber{6} \acmArticle{CSCW173}
\acmMonth{10} \acmDOI{10.1145/3817021}

\begin{document}
\title[Third-Party Developers' Support Seeking and Provision]{Twitch Third-Party Developers' Support Seeking and Provision Practices on Discord}

\author{Jie Cai}
\affiliation{%
  \institution{Tsinghua University}
%   \streetaddress{1 Th{\o}rv{\"a}ld Circle}
  \city{Beijing}
  \country{China}}
\email{jie-cai@tsinghua.edu.cn}
\orcid{0000-0002-0582-555X}

\author{He Zhang}
\affiliation{%
  \institution{Pennsylvania State University}
%   \streetaddress{1 Th{\o}rv{\"a}ld Circle}
  \city{University Park}
  \country{USA}}
\email{hpz5211@psu.edu}
\orcid{0000-0002-8169-1653}

\author{Yueyan Liu}
\affiliation{%
  \institution{Tsinghua University}
%   \streetaddress{1 Th{\o}rv{\"a}ld Circle}
  \city{Beijing}
  \country{China}}
\email{yueyan_liu@mail.tsinghua.edu.cn}
\orcid{0009-0002-8591-1374}

\author{John M. Carroll}
\affiliation{%
  \institution{Pennsylvania State University}
%   \streetaddress{1 Th{\o}rv{\"a}ld Circle}
  \city{University Park}
  \country{USA}}
\email{jmc56@psu.edu}
\orcid{0000-0001-5189-337X}

\author{Chun Yu}
\authornote{Corresponding author.}
\affiliation{%
  \institution{Tsinghua University}
  \city{Beijing}
  \country{China}
}
\email{chunyu@tsinghua.edu.cn}
\orcid{0000-0003-2591-7993}

\renewcommand{\shortauthors}{Cai et al.}

\begin{abstract}
Third-party developers (TPDs) often turn to online communities for support when they can't get immediate responses from the platform. Twitch, as a leading live streaming platform, attracted many TPDs and formed an online support community on Discord. This study explores TPDs' support practices via mixed method (a topic modeling to identify topics related to support seeking and provision first and a follow-up in-depth qualitative analysis with these topics) and found that: (1) TPDs' support-seeking practices around social, technical, and policy matters are highly dependent on Twitch, and this dependence acts as a form of platform labor; (2) TPDs need to switch between Discord and Twitch regarding seeking and provision, exacerbating TPDs' platform labor; (3) TPDs' flexible role practices reflect the community's flourishing on Discord but require roles to bridge the two platforms and transfer informal support seeking to possible formal support from Twitch. We propose implications for effectively managing support seeking and provision between formal and informal spaces to improve the development of TPDs. We also contribute to community support practice and to platform ecology work in CSCW. 
% \\
% \textcolor{red}{Preprint accepted at ACM CSCW 2026}
\end{abstract}




\received{May 13, 2025}
\received[revised]{January 13, 2026}
\received[accepted]{April 9, 2026}
% \newpage

\maketitle

% \textcolor{pink}{Prepint accepted by ACM CHI 2024}

\section{Introduction}

As a prominent live-streaming platform, Twitch\footnote{\url{https://www.twitch.tv/}} attracts a large number of users and delivers diverse streaming content, including games, music, creative content, and e-sports. To support the live streaming platform, third-party developers (TPDs) have leveraged Twitch application programmers' interfaces (APIs) and contracts, and developed numerous tools (e.g., bots, extensions, plugins) to enhance streamer-viewer interactions and facilitate community management \cite{han_hate_2023, cai_hate_2023, cai_categorizing_2019}. Nonetheless, TPDs often encounter challenges in their iterative development process, such as adapting to frequent technological updates and interpreting vague platform guidelines \cite{cai_third-party_2024}. The insufficient feedback and delayed responses prompt TPDs to seek informal support from other online communities.

Prior CSCW  research on developers' support has explored broad contexts, such as open-source software communities \cite{hellman_ohhh_2025}, indie or pro-amateur game development \cite{freeman_pro-amateur-driven_2020, freeman_mitigating_2020}, from diverse perspectives, such as participatory innovation \cite{freeman_innovation_2019}, social capital \cite{terry_perceptions_2010}, and the barriers to community contribution \cite{steinmacher_overcoming_2019}. However, these studies typically focus on professional developers with formal or institutional support \cite{freeman_exploring_2019}. Unlike professional developers, TPDs face many challenges, such as unclear platform rules, unpredictable APIs and policy changes, and technical barriers \cite{cai_third-party_2024}, and lack a clear support system.

Therefore, understanding TPD's support mechanisms is crucial for understanding how TPDs sustain innovation and community participation under limited formal support, and for revealing the broader implications for platform ecosystems. This study addresses the following research questions:

\begin{itemize}
    \item What support do Twitch TPDs seek from the Discord community?
    \item What roles do Twitch TPDs play to provide support in the Discord community?
\end{itemize}

This study aimed to understand TPDs' support practices and identify design opportunities to facilitate them. We collected and analyzed discussion content from a Discord community using topic modeling and in-depth qualitative analysis. We found that TPDs play multiple, flexible roles on Discord. However, they depend highly on seeking support and obtaining interpretations of policies and technical problems. We revealed the social dynamics within the Discord community and analyzed the interaction dynamics among roles, support-seeking, and provision.

\section{Background: Discord Community for Twitch Third-Party Developers}

We chose Twitch as the research context for two reasons. On the one hand, the live streaming industry, with its strong interactivity and multimodal integration \cite{hamilton_streaming_2014}, maintains a high-speed developing trend driven by technological iterations. On the other hand, as a leading platform in the industry, its user groups, such as streamers \cite{wohn_audience_2020}, viewers \cite{seering_shaping_2017, yen_storychat_2023}, and volunteer moderators on Twitch have become research hotspots in HCI (e.g., \cite{cai_coordination_2022, cai_understanding_2023}, etc.). However, it is worth noting that TPDs who develop dedicated tools for these groups have often been overlooked in research.

\begin{figure*}
  \centering
  \includegraphics[width=0.9\linewidth]{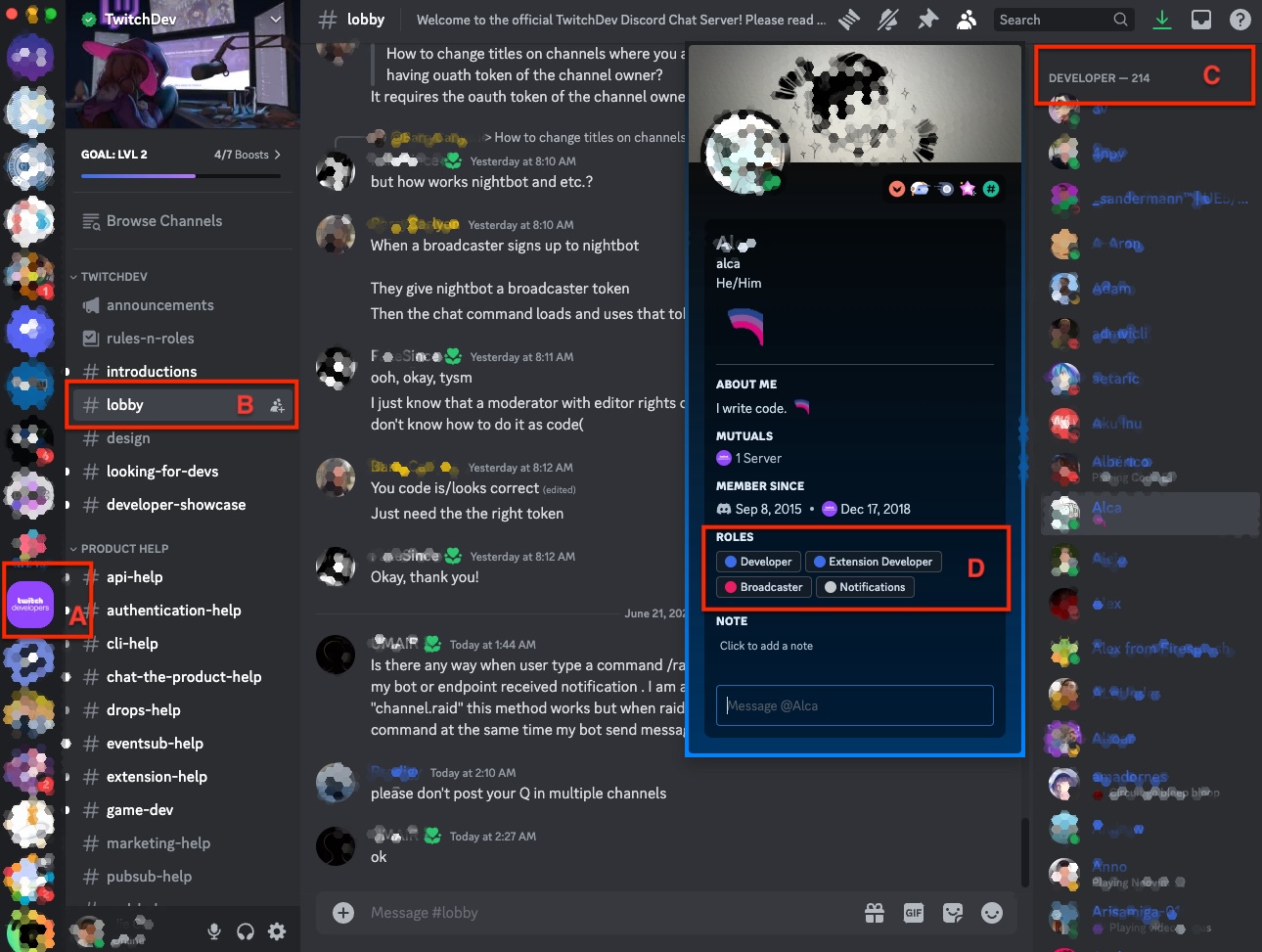}
  \caption{A screenshot of the \textit{TwitchDev} server on Discord. (A) the server icon and name; (B) the lobby channel as the main channel for discussion; (C) a list of 214 active developers online; (D) a user with many different tagged roles verified by the server administrators, such as \textbf{developer} and \textbf{extension developer}.}
  \Description{A screenshot of the Twitch developer server on Discord. (A) the list of 193 active developers online; (B) a user with many different tagged roles verified by the server administrators, such as \textbf{developer} and \textbf{extension developer}; (C) various channels for developers to discuss and collaborate on technical issues related to bot development.}
  \label{discord}
\end{figure*}

Twitch launched its Extension program in 2017 to delegate development tasks to TPDs to meet the diverse needs of streamer community management \cite{gartenberg_twitch_2017}. Third-party tools on Twitch, including extensions and bots, enhance live streams by fostering interaction between streamers and viewers. Twitch TPDs are driven by skill development and career growth opportunities. TPDs were not directly compensated by Twitch; instead, by various methods to work with streamers and viewers, such as developing extensions that allow viewers to spend Bits -- a virtual currency related to money spending,  and share the income with the streamers (e.g.,  20\% cut of Bits spent in the tool), subscriptions from streamers and viewers for premium bot features, and donations from the community.
The platform provides APIs and resources to facilitate collaboration among TPDs and other developers, while also benefiting from the outcomes of their contributions \cite{durmus_twitch_2024}. Sometimes, the platform even captured TPDs' products for its own features, potentially sidelining TPDs in favor of competition against the platform (e.g., Third-party donation management tools vs. Twitch Bits, the platform's first-party donation management tool) \cite{partin_bit_2020}. This dependence and power imbalance can create tension and precarity, negatively impacting the platform ecosystem. 
Discord plays a unique role in the Twitch ecosystem. Discord was initially used as a voice collaboration tool for gaming streamers to communicate during team play, but it has gradually evolved into an essential platform for daily social interaction between streamers and their viewers. Now, it has become a gathering place for interest groups \cite{hwang_adopting_2024,jiang_moderation_2019}. Due to the integration between Twitch and Discord, many TPDs have formed professional communities on Discord. The `TwitchDev' server, to which our research team belongs, is a typical example; this community is operated by volunteers. TPDs only need to prove that their projects serve Twitch users or use the platform's open interfaces to obtain community developer badges. Currently, this community brings together thousands of active developers, with an average daily participation of over 2,000. The members include formal Twitch staff, TPDs, streamers, and viewers. A Discord server has many channels; the `Lobby' channel, as the core communication area, serves multiple functions, including information release, technical Q\&A, and social icebreakers. The main interface with features is shown in \autoref{discord}. Each member can be assigned multiple roles. In the \autoref{discord}, the member had four different roles: Developer, Extension Developer, Broadcaster, and Notifications. According to the `rules-n-roles' channel, there are at least other roles, such as Game Developers and Designers. We also observed some event roles and other roles, such as `Twitch Staff'. In this study, we used TPD as an umbrella term to encompass various roles that are not designated as Twitch Staff but rather tagged with `developer' in the roles, such as Developers, Extension Developers, and Game Developers. These roles are not explicitly associated with Twitch, but they contribute to the platform's ecosystem with third-party tool development.

\section{Related Work}
\subsection{Third-Party Developers and Support}
TPDs develop applications, services, and systems that meet end users' needs within the platform's architecture \cite{ghazawneh_balancing_2013}. The partnership between TPDs and platforms is usually clearly defined through contracts \cite{boudreau_how_2009}. This contractual relationship replaces the traditional salary system, enabling developers to reach a broader user market through the platform \cite{west_browsing_2010}. The two parties formed a mutually beneficial and symbiotic innovation collaboration mechanism. With their novel ideas and technical capabilities, TPDs have enriched the platform ecosystem's diversity. On the other hand, platforms leverage these innovations to expand service offerings, enhance their business competitiveness, and drive value growth \cite{qi_understanding_2021}. This collaborative model also brings significant operational advantages to the platform: the platform does not need to invest a large amount of resources to form a huge internal development team, effectively reducing labor and resource costs; on the other hand, by integrating the crowdsource of the developer community, it can quickly respond to market changes, cultivate a vibrant digital ecosystem \cite{boudreau_let_2012, ghazawneh_balancing_2013}.

The existing HCI literature on developers mainly focuses on game developers \cite{freeman_mitigating_2020,freeman_understanding_2023, freeman_pro-amateur-driven_2020, li_exploring_2023}, open-source software development \cite{huang_effectiveness_2016, sanei_untold_2025, gutwin_group_2004}, and hardware development \cite{mellis_collaboration_2012}, with topics specifically focusing on supporting collaboration \cite{guzzi_supporting_2015, freeman_making_2016, zhang_who_2025}, teamwork \cite{freeman_exploring_2019,freeman_tale_2021}, and conflicts \cite{filippova_mudslinging_2015}. For example, TPDs face unique challenges in their development practices, especially when dealing with the platform's technical updates and multi-layered policy regulations \cite{cai_third-party_2024}. This complexity and lack of knowledge may lead novice developers to reuse existing resources with a limited understanding of their inner workings \cite{yang_grounding_2018}. Online support provides empirical experiences from peers, which complements knowledge from professionals and offers additional access to formal knowledge \cite{gui_investigating_2017}. It is necessary to understand the situation of TPDs within the platform and to explore their multi-dimensional support needs to improve the platform ecosystem.

\subsection{Support Seeking and Provision in Online Community in HCI}

Social support involves the exchange of resources between individuals, with either the provider or the receiver explicitly aiming to enhance the receiver's well-being \cite{shumaker_toward_1984}. This concept serves as a lens for understanding supportive behaviors within online communities \cite{nick_online_2018,oh_how_2014,barta_similar_2023}. Prior work in HCI and CSCW has discussed social support in various online communities, including online health communities \cite{skeels_catalyzing_2010}, and E-sports communities \cite{freeman_exploring_2019}, and has explored diverse user groups, such as users of different genders in social VR \cite{li_we_2023,freeman_tale_2021}, streamers in live streaming \cite{wohn_understanding_2015}, listeners in social music platforms \cite{jin_understanding_2023}, and marginalized groups in different online communities (e.g., Facebook, Reddit, Instagram etc.) \cite{zhou_veteran_2022,gui_investigating_2017, yao_together_2021}, for health information support \cite{chancellor_norms_2018,wang_cass_2021, sharma_mental_2018}.  

Support practices can be formal and informal \cite{french_formal_2020}; both are important to the individual who seeks and provides support \cite{gouin_impact_2016}. Formal support structures encompass organized resources such as counseling services, support groups, and structured programs within institutions or local communities \cite{prost_analysing_2010}. These mechanisms are typically established to provide professional guidance, structured interventions, and specialized assistance tailored to specific needs \cite{wohn_understanding_2015}. Formal support in the online space includes the platform's professional support service, official documents, and guidelines. On the other hand, informal support in online spaces, such as simply participating in social acts of meaning-making through conversations within social networks, can improve their coping process when experiencing difficulties. For example, informal support can be observed in the form of online comments, virtual hugs, or affirmations that validate individuals' experiences and emotions \cite{li_we_2023}. 
The lack of formal support from the core space might drive users to peripheral spaces for informal support to fulfill their needs and commits \cite{gui_investigating_2017}. 

Informal support in online communities is also widely discussed in HCI, including diverse support-seeking and provision behaviors and the reasons behind them (e.g., \cite{wohn_explaining_2018, kim_receivers_2022, johnson_its_2022}).
Community members involved in diverse roles, 
such as experts, support seekers, and storytellers \cite{yang_seekers_2019, singh_types_2011, haraty_online_2017}.
Most informal support is voluntary, which means community members provide additional labor and may be invisible (hard to quantify) to the core space they are working on \cite{li_all_2022, li_measuring_2022}. Noteably, the labor related to the informal format happened in another platform is different from the mainstream labor discussion about gigwork around crowdsource workers (e.g., Amazon Mechanical Turk \cite{irani_turkopticon_2013}), ridesharing drivers (e.g., Uber and Lfyt \cite{gloss_designing_2016, lee_working_2015}), beauty workers \cite{anjali_anwar_watched_2021}, freelancers (e.g., Upwork \cite{kim_gender_2025, huang_design_2024}, and content creators (e.g., YouTube \cite{ma_conceptualizing_2023}) in HCI and CSCW as these groups are highly dependent on the platform where the labor actually happened, and the platforms can manage the labor with algorithms for profit \cite{tarafdar_algorithms_2023, stelmaszak_when_2025}. 
These attributes may make TPDs' labor experiences different from those of other workers. 

TPDs can seek support from their offline networks and receive professional, formal support from their peers, etc. However, we know very little about how they seek and provide support informally among people with weak ties (mostly strangers)
in an online space different from their serving platform. We extend this line of research to focus on Twitch TPDs' informal support practices within the Discord community, exploring the diverse roles they play and the labor they take in the support-seeking and provision process. 

\section{Methods}
We applied a mixed-methods approach to explore TPDs' social support practices through their natural conversations in an online community, using data from their online interactions. The first and second authors are active Discord users who periodically engage with the \textit{TwitchDev} community and often observe TPDs' interactions.
We first crawled all the chat history of the Discord server. After we went through all the channels, we realized that the `lobby' contains the most extensive dataset. All other channels have some issues according to our observation: (1) the other channels are compartively small and extended or being fowarded from the `lobby' channel, where the main topics are covered; (2) the most are temporal for specific topics, which might be deleted later and hard to track the conversation; (3) some are private for personal or sensitive topics without granted permissions. Then, we decided to focus on the `lobby' as the primary data source. The topic modeling is to identify the main trends TPDs discussed in the large data corpus, with two purposes: (1) to have an overview of the data, and (2) to identify social support-related topics with a comparatively condensed data corpus for further analysis. After identifying the relevant topics, we conducted a thematic analysis to gain an in-depth understanding of their support practices.

\subsection{Data Exploration and Preprocessing}
\label{sec:data-exploration}

We gathered a dataset comprising 45,376 messages from the `lobby' channel, the most active channel. Each entry contains the message content and its corresponding timestamp. To prepare the data for analysis, we employed the NLTK library\footnote{\url{https://www.nltk.org/}} to remove English stopwords from all messages. Drawing on established practices in text analysis~\cite{tseng2007text} and informed by prior research with similar data structures~\cite{ma_how_2021}, we implemented a quality filter to exclude messages containing fewer than ten words after stopword removal. This preprocessing step helps eliminate short, non-informative texts, ensuring our analysis captures meaningful contextual information.

After preprocessing, our final dataset consisted of 8,219 data entries containing 279,581 words (1,567,282 characters). We clarified that our data collection represented a single, fixed extraction from the channel, and we did not account for channel activity or changes that occurred before or after this collection period, as this was in accordance with our research questions and practical constraints. This data history spans from October 25, 2018, the date of the first message in the `lobby', to April 25, 2023, the date we initiated this project. We acknowledged that the limitation of data collection, as new issues could arise in discussions after that time point, but we focused not on the context but on the high-level challenges of support seeking and provision. 

In this study, we examined only publicly available comments and did not engage with human subjects, so it is exempt from the university's ethical research review board. To protect user privacy, we followed the \textit{Internet Research: Ethical Guidelines 3.0} \cite{franzke_internet_2020} and anonymized the dataset by removing identifying details (e.g., user IDs and usernames). We reported only the comment text and replaced any usernames with @XXX if someone was mentioned. 

\subsubsection{Topic Modeling}\label{subsec:topicmodeling}
Topic modeling is a widely used method for uncovering patterns and themes in large collections of text. Unlike approaches that rely solely on keyword retrieval, topic modeling helps discover hidden thematic structures by modeling the probabilistic distribution of words across topics~\cite{10.1145/2133806.2133826}. This method is particularly helpful when exploring posts and comments on the Discord platform, which typically consist of short texts characterized by sparsity, noise, and non-informative data, and has been widely applied in similar contexts~\cite{likhitha_detailed_2019}.

In this work, we employed Latent Dirichlet Allocation (LDA)~\cite{blei_latent_2003} for topic modeling. The LDA parameters Alpha and Beta were set to default values of 1.0 divided by the number of topics. We conducted a cluster analysis on the preprocessed data and identified an optimal number of clusters. This analysis informed our selection of eight topic trends.

To validate and complement our observations from the hierarchical clustering analysis, we further employed coherence scoring as an independent evaluation method. Coherence scores are widely regarded as among the most effective metrics for assessing topic modeling quality~\cite {10.5555/2390948.2391052}, providing a quantitative measure of semantic consistency within identified topics. Following the methodology outlined by Stevens et al.~\cite{stevens_exploring_2012}, we systematically evaluated coherence scores across a range of potential topic configurations (7-20 topics). The results corroborated our initial clustering insights, with the coherence metric peaking at eight topics, thus confirming this as the optimal number for our analysis.

\autoref{tab:LDAcorpus} illustrates the computational extraction of topic distribution in our results, generated using LDAvis~\cite{sievert_ldavis_2014}. This approach enables each comment to be generated from multiple topics, with each topic comprising multiple keywords that characterize its thematic content.

\begin{table*}[t]
\caption{The results of the LDA topic modeling analysis on the corpus. Each topic is represented by weighted terms (the top 10 highest). The subtopics we focus on are highlighted in red, and the topics we choose for further in-depth analysis are highlighted in gray.}
\resizebox{\textwidth}{!}{%
\begin{tabular}{l|rrrrrrrrrr}
\hline
 & Term 1 & Term 2 & Term 3 & Term 4 & Term 5 & Term 6 & Term 7 & Term 8 & Term 9 & Term 10 \\ \hline
\rowcolor[HTML]{EFEFEF} 
Topic 0: & 0.043*"twitch" & {\color[HTML]{CB0000}0.025*"help"} & 0.020*"discord" & 0.019*"party" & 0.014*"com" & {\color[HTML]{CB0000}0.010*"support"} & 0.010*"account" & 0.010*"new" & 0.010*"third" & 0.010*"report" \\
Topic 1: & 0.022*"channel" & 0.017*"still" & 0.013*"would" & 0.013*"want" & 0.011*"people" & 0.010*"well" & 0.010*"name" & 0.009*"stream" & 0.009*"since" & 0.009*"channels" \\
Topic 2: & 0.022*"game" & 0.017*"twitch" & 0.012*"time" & 0.012*"see" & 0.012*"use" & 0.011*"example" & 0.010*"work" & 0.009*"thing" & 0.009*"uservoice" & 0.009*"guess" \\
\rowcolor[HTML]{EFEFEF} 
Topic 3: & 0.039*"thanks" & {\color[HTML]{CB0000}0.030*"answer"} & 0.023*"http" & 0.020*"quest" & 0.015*"got" & {\color[HTML]{CB0000}0.014*"answers"} & 0.014*"join" & 0.012*"barry" & {\color[HTML]{CB0000}0.010*"help"} & 0.009*"language" \\
Topic 4: & 0.025*"extension" & 0.019*"chat" & 0.015*"twitch" & 0.015*"good" & 0.012*"message" & 0.011*"use" & 0.009*"code" & 0.009*"token" & 0.009*"need" & 0.009*"bits" \\
\rowcolor[HTML]{EFEFEF} 
Topic 5: & 0.072*"twitch" & 0.030*"tv" & 0.019*"api" & 0.016*"contact" & {\color[HTML]{CB0000}0.015*"ask"} & 0.015*"dev" & 0.015*"user" & 0.012*"send" & 0.012*"luck" & 0.012*"know" \\
Topic 6: & 0.018*"one" & 0.017*"use" & 0.016*"something" & 0.015*"thank" & 0.011*"already" &  0.010*"bots" & 0.010*"wait" & 0.010*"go" & 0.009*"looking" & 0.009*"think" \\
\rowcolor[HTML]{EFEFEF} 
Topic 7 & 0.033*"twitch" & {\color[HTML]{CB0000}0.031*"support"} & 0.030*"link" & 0.026*"us" & 0.022*"found" & 0.021*"looks" & {\color[HTML]{CB0000}0.021*"help"} & 0.019*"request" & 0.015*"please" & 0.015*"friends" \\ \hline
\end{tabular}%
}

\label{tab:LDAcorpus}
\end{table*}

\subsection{Thematic Analysis} 
In this study, we investigated TPD's support practices, focusing on key terms associated with support seeking and provision, including ``ask,'' ``help,'' ``answer,'' and ``support.'' We chose these words because they are explicitly related to support seeking and provision. We identified four specific topics relevant to our focus: Topic 0 (help and support), Topic 3 (answer and help), Topic 5 (ask), and Topic 7 (support and help). These topics are closely related to support seeking and provision, while other topics primarily addressed general information, which was also covered by other topics. Therefore, we concentrate solely on these four topics. After confirming the four topics, we organize the comments chronologically to create a sequence of conversations that provide context for our thematic analysis. Though some short data entries have been removed, the sequence of conversations allows us to grasp the main discussion timelines, topics, and structures, which will aid our support-seeking and provision analysis.

To analyze these comments, we employed thematic analysis \cite{braun_using_2006} with both inductive and deductive coding methods \cite{fereday_demonstrating_2006}. First, all researchers went through the condensed data corpus to familiarize themselves with the TPDs' support-seeking and provision practices. Next, two researchers independently coded a sample of 200 randomly selected comments. After a calibration meeting, they refine their coding to ensure consistency, ultimately developing an initial codebook with 52 codes, plus two functional codes: 0 (irrelevant) and 53 (relevant but not listed). We added these two functional codes because the datasets contain comments from streamers and viewers. If the comments were not specifically related to development practices or were too short to indicate that they were from TPDs clearly, the two researchers marked them as irrelevant. If the two researchers felt the comments were interesting/relevant but not covered by the existing codebook, they marked them as relevant but not listed. 

They then apply this initial codebook to an additional random sample of 200 comments to test its validity. This process yields a Cohen's kappa of 0.72, indicating acceptable agreement given the size of the codebook. The researchers discuss and resolve any inconsistent codings, which lead to individual coding efforts. During this process, they meet weekly with a senior researcher to review the progress and address discrepancies.

After completing the coding process for all data entries across the four selected topics  (totaling 3,375 data entries, 111,515 words, and 530,123 characters), we transferred all identified codes to a Miro\footnote{\url{https://miro.com/index/}} whiteboard. The three researchers collaboratively organize the codes into subthemes and themes that answer the research questions. The code ``0'', including 117 data entries, was not in the final theme categorization. Since the reliability is .72, some quotes slip in that are not considered from TPDs, so three researchers cross-validated by checking the quotes and codes in the categorization and the following writing. Finally, we compile relevant quotes associated with each code into a shared document to draft paragraphs, topics, and themes, ensuring we provide appropriate justification for our findings.

\section{Findings}
\subsection{Types of Support Twitch TPDs Seek In the Discord Community}
TPDs sought various forms of support on the Discord server, ranging from inquiries about the Twitch platform and its support system, including complaints, to technical development questions and policy and guidance, as well as socializing activities, such as learning and hardware inquiries.

\subsubsection{Inquiry of Account Registration and Verification}
TPDs asked questions about their account registration and verification. Many developers were asking about how to change their mobile numbers. One TPD asked, \say{\textit{Hello I am waiting for the change of my number from my twitch account already sent the same email 3 times / We are reviewing your request to change your phone number 2FA, you know if you have to wait?}} This developer wondered how long someone would have to wait for their phone number to be changed for two-factor authentication (2FA) purposes. Because they couldn't reach Twitch support, they asked their fellow developers for help gathering information. 

Some TPDs also faced registration issues on Twitch. A TPD asked, \say{\textit{Greetings! I can't register my company on twitch. My request just disappeared after some time. Can anyone aid me? Maybe someone from twitch staff here?}} This TPD is having trouble registering their company on the Twitch site and is looking for assistance from the Discord server, and even mentioned the Twitch staff, who also monitored the community.

\subsubsection{Inquiry of Twitch Support Speed}

TPDs often discussed their trouble communicating with the Twitch platform. Many TPDs reported slow or no response from the platform. In fact, it was very common for TPDs to wait weeks after submitting a support request before receiving a response from Twitch. 
One of the most common complaints about TPD was the lengthy wait for a response from Twitch support or staff. For example, a TPD complained, 

\begin{quote}
I have wrote a request to support, but so far no reply in about 4 days. That sucks frankly, because my sister is already building up a fan base, people are enjoying the fact she is funny and they find her attractive. She is very flattered by that and wants to grow on the Twitch platform and do a huge amount of work. If anyone can help us, that would be amazing! Thanks in advance for the help! With all due respect!
\end{quote}
In this instance, the TPD submitted a support request so that the TPD's sister could better manage her Twitch platform. The TPD was frustrated by the lack of acknowledgment that their request had been completed and was concerned that it would continue to take a long time.

\subsubsection{Complaint of Twitch's Support System}
Sometimes, TPDs expressed that Twitch's support system failed to help them in other ways, aside from responding too slowly. A TPD said, \say{\textit{Twitch removed all formal developer relation contact channels. The only one if the main support form right now.}} Here, the TPD noticed that Twitch removed other forms of possible support for developers, which limited their chances of getting help. Another TPD found that there were specific types of questions Twitch support would not answer, relating to recycling: \say{\textit{always got the standard answer that twitch doesn't answer or give informations about the recycle process.}} Finally, if there was an answer from Twitch, it was likely an error message that provided little helpful information. This lack of support from Twitch indicated disorganization in their support request-handling system.

Other TPDs complained about communication beyond support requests. A TPD recounted their struggle to obtain information on changing a username and password after Twitch suffered a security breach: \say{\textit{Can anyone help me out with this...So u all might know that twitch suffered a major breach And ik that u have to change password and stuff but i when i request for a password reset email...it doesn't respond.}} The TPD expressed their concern and asked other TPDs for help, since they received no response from the Twitch platform. Such a situation revealed potential challenges that TPDs might face in the event of a security breach or internal conflicts within Twitch.

TPDs had numerous questions regarding Twitch emails and accessing Twitch support via email. These responses were often developed in response to a lack of response from Twitch's support ticket system. A TPD asked, \say{\textit{hey guys is there a support mail from twitch to make them change my company based mail address.}}  By asking the fellow developers if they knew how to contact Twitch,  they would possibly find information faster than through Twitch directly. Other messages related to emails still expressed impatience regarding the wait time for responses. Another TPD questioned how much longer their email had token to be answered since they were associated with an organization, inquiring, \say{\textit{Any idea when mails regarding organization questions are being answered? Waiting to get help to claim this and we have a team waiting.}} Because emails were a common way for Twitch to get into contact with its users, it was unexpected that Twitch users would have a difficult time reaching the platform via that method.

\subsubsection{Technical Development Questions}

TPDs asked other TPDs about their technology development, including third-party tools (bot design and detection), extension setup and management, and databases and Twitch integration.

Many TPDs had questions about bots, specifically how to develop them and prevent bot spam. A TPD asked, \say{\textit{Hello, where can I ask for some assistance related to Java (I need to use that) libraries to design an IRC bot?}} Another TPD asked about design solutions to prevent bot spamming, as they were concerned about systemized spam attacks. TPDs also have questions about extension support, specifically how to build Twitch extensions. A TPD asked where to find the code for their front-end extension and how to attach an extension in its unpublished state to their Twitch channel:
\begin{quote}
How do I tell Twitch where to find the code \& assets for my front-end that will live in the provided IFRAME? Also, how do I attach my Extension while it is in an unpublished state to my Twitch channel so I can see it in my Extensions list? Do I have to be a Twitch Prime member to do that?
\end{quote}

TPDs also had questions regarding the use of the Twitch extension on mobile systems:
\begin{quote}
   Hello, can someone help? Our team is planning to develop a twitch extension, and I wonder do twitch extensions work in mobile browsers? Not Twitch app, but Safari or Chrome. Would be grateful for any help and advices.
\end{quote}
Similarly, another TPD asked, \say{\textit{I'm trying to embed the twitch chat in a website and it seems to work fine on desktop but not on mobile.. anybody know about this?}}

TPDs asked questions about databases and Steam, a platform with many video games. A TPD, who was also tagged as a broadcaster and developer, asked: \say{\textit{Anyone here used MongoDB before? Is it more logical to use one database for users and another for my other data, or should I only use one database where I store the data separated by collections?}} This TPD was confused about whether or not to use one or multiple databases for other users and themselves. Questions about running the frontend were also frequently asked by TPDs.

\subsubsection{Inquiry about Policy and Guidance}
TPDs asked about legal advice. There could be multiple legal topics a TPD could ask about, such as clarification on a legal matter or copyright questions. The TPDs usually did so to prevent themselves from breaking Twitch or Discord's rules, or, in more severe cases, potentially breaking the law.
For example, several TPDs who held multiple roles in the community discussed legal matters.

\begin{quote}
TPD\_A: Hey, I have received a DMCA copyright infringement complaint regarding using Twitch thumbnails (we only use the available data from Twitch's formal API). They are going to take down the website in 24hours if we do not delete it, is there an email I can contact Twitch with since there's no option in the contact support form specific to this issue?

TPD\_B (extension developer and broadcaster): Who sent you the email? Twitch isn't the copyright holder, so they wouldn't have anything to do with it.

TPD\_A: @TPD\_B a lawfirm contacted cloudflare, cloudflare contacted my cloud provider and they have contacted me.

TPD\_C (extension developer and broadcaster): @TPD\_A  legal@twitch.tv is your best bet. I don't understand what they're claiming though. Just block their account from showing up.

\end{quote}
This conversation among TPDs showed how they seek legal advice and provide appropriate ways to handle the situation.

\subsubsection{Learning Inquiry}
TPDs would announce in the chat that they were novice developers or wanted to start learning a new development topic. After announcing this, they would usually follow up with a question on how Twitch started or a specific question for other developers. In this way, TPDs were able to initiate the conversation, clarify their level of knowledge, and receive the best advice applicable to their situation. 

For example, a TPD commented, \say{\textit{I'll have to get learning React. Small steps = big steps later on,}} making their intentions known in the chat, and they might follow up with some questions on how to get started or a specific problem they were facing as a beginner:

\begin{quote}
Hey y'all. I'm trying to figure out how to create an application that is integrated with my twitch channel, but I can't find any learning material for complete beginners. Is there something that can help me figure out how to use the twitch API?
\end{quote}
In this case, the TPD, also a broadcaster, specified the type of help they needed, and by stating they were a beginner or looking for beginner material, they received the proper advice from other TPDs.

\subsubsection{Inquiry of Hardware}
TPDs asked questions about hardware recommendations for entertainment in addition to development, and other TPDs provided recommendations. Questions typically concerned hardware performance, and the results would directly address them. A TPD says, 
\begin{quote}
Anyone have a recommendation for a GPU that can play games such as OW / csgo on low / med settings? my GPU died a week or two ago and want to replace it, i had a 480 for reference but that was overkill for the games i played.
\end{quote}
In response, another TPD stated, \say{\textit{rx 580 would do ok or go for 1060 6gb or at least 3gb, or just get 1660 ti if you like Nvidia and their color settings and such.}} Here, TPDs were conversing about the best GPU for a certain gaming condition. Another TPD made an offhand comment regarding the games they play and GPUs that were needed for it, \say{\textit{haha yeah, i don't really play that many games and the games i do play don't really require a high end gpu.}}

\subsection{Type of Roles Twitch TPDs Play to Provide Support in the Discord Community}
TPDs played diverse roles to provide support to others; they can be instructors to show step-by-step processes, explainers to provide reasons, advisors to give tips to handle situations, support redirectioners to share resources in other communities, socializers to share thoughts and organize gatherings, conflict facilitators to keep civil discussions, and complimenters to show appreciation and applaud others.

\subsubsection{Instructor}
TPs provided step-by-step instructions for completing a specific task, rather than offering advice or general explanations, including guidance on requesting help from Twitch and registering for an app. Experienced TPDs who were more knowledgeable in the specific field attempted to assist confused and novice TPDs.

Instructions about app registration were commonly explained by TPDs around the server: \say{\textit{The ONLY way you'll get help in these matters is to go through the proper methods, such as the ticket system, so that you can be put in a queue, wait, and properly processed.}} This TPD explained how other TPDs would have to go through the correct and legal methods to get help in some way, such as getting registered for the app.

Another topic related to instructions was providing suggestions. For example, a TPD provided the server with links to CSS frameworks, which are libraries of code that help implement website layouts. However, this TPD also made a CSS Framework as a personal project, so they were explaining it to the server. 

\subsubsection{Explainer}
TPDs explained the Discord community boundaries and limitations to support seekers. There were two main explanations for TPDs. First, a TPD could explain the API and the server's API specialties. Second, TPDs could explain that the Discord server is a third-party server and that it was not typically used to contact Twitch staff or to seek help with a specific problem that only Twitch staff could resolve.

Many TPDs are curious about API and how the server's TPDs are interested, so the API TPDs explain. A TPD said, \say{\textit{It is the formal Developer server in that a lot of the API peeps are here and help as well as take API feedback.}} However, the server members only know so much about the API. One of the TPDs confirmed this by saying: \say{\textit{This being a community dev discord we only know so much. And generally only do help on third party product stuff like API and chat.}}

Some TPDs specified channels where TPDs could post about API. A TPD explained, \say{\textit{Thats why I asked where… You posted in \#api-help and \#authentication-help which is not relevant to your question where \#extension-help is, so I needed to know "where" you meant.}} This TPD explained the different API-related channels on the server and asked another TPD about their API, authentication, and extension-related questions.

Other TPDs explained that the server was a third-party server, not a formal Twitch platform server. Some TPDs had to explain that they couldn't help in certain circumstances because of the nature of the server: \say{\textit{Use that. That will go to the people who can help you. As i told you earlier this is a 3rd party developer community and we can't help you with your issue.}} TPDs also explained how sometimes Twitch staff came by the Discord server, but talking to Twitch staff was not the main purpose of the server. They continued by explaining that if other TPDs had ideas to improve Twitch, they could post them on Twitch's UserVoice page, a website dedicated to collecting community feedback. A TPD explained what the server was to new TPDs in the server: 
\begin{quote}
    Welcome \o This Discord server is dedicated to 3rd party developers for Twitch tools. For example, Extension developers, game developers, chatbot developers, etc. People can get technical help for their projects here on various topics.
\end{quote} 
The TPD explained that the server was for all kinds of third-party developers on Twitch. 

TPDs explained coding and certain aspects of code. The primary explanations relate to React, the front-end library Twitch uses, and Redux, the back-end library Twitch uses. TPDs explained these concepts to other TPDs to help them learn React and Redux and how they worked with Twitch:
\say{\textit{If I'm correct React is the Front end and I believe Redux is the back end but I've not heard of GraphQL before.}} This TPD said that React was the front-end (the part of the code that the user of the program was supposed to see), while Redux was the back-end (the part of the code where the user of the program was not supposed to see). However, this TPD has never heard of GraphQL before.
Another TPD added, \say{\textit{And as Twitch use React, would make sense to show that off for people who are considering looking into making more hefty apps.}} 
Another TPD further explained that Twitch was primarily a React-based company, which helped other TPDs understand their code.

\subsubsection{Advisor}
Advisors differed from instructors or explainers as they offered tips rather than step-by-step instructions or explanations. TPDs shared advice and information on specific topics, such as devices or the rationale for what to do or avoid, with the aim of helping inexperienced TPDs with insights into development and Twitch operations.

Many senior TPDs had extensive experience with Twitch, as well as GitHub, a website and programming platform where TPDs can save their code for themselves or others to share. A TPD said, \say{\textit{Pitfalls for beginners: There is no git repo or something that can you clone. Every tutorial is a `.streamDeckPlugin' file (zipped) that also contains the source. at the moment.}} Which explains to novice TPDs how there's no git repository for them to copy code from for their Twitch projects.

TPDs also gave tips and recommendations to peers. For example, a TPD who is also a broadcaster gave tips about screen quality to maintain potential viewers: 
\begin{quote}
    Affiliate can't rely on always having access to transcoding so I'd recommend not using 1080p60 anyway as when you don't have transcoding you could be losing a large part of potential viewership as not only the demand of 1080p60 is beyond some users, there's also the bandwidth requirements due to requiring a higher bitrate (if you don't want it to look awful).
\end{quote} 
The TPD noted transcoding wasn't always available, so they advised avoiding \say{1080p60} to reduce bandwidth demands and prevent lag or quality issues.

\subsubsection{Support Redirectioner}
Twitch TPDs often sought support on this Discord server when they cannot reach Twitch Support or cannot find answers elsewhere. Other TPDs often redirected them to official support links (sometimes via automated messages), including suggestions for alternative Discord servers or websites that might be better suited to the question.

One of the most common messages sent on the server occurred after a TPD typed something in chat that indicated they needed help from Twitch support. If a TPD triggered this system, the following message would be sent to the server:  \say{\textit{Hey there! Looks like you found us on a quest for answers. You're in luck. Our friends over at TwitchSupport have the answer for you. You can contact them at http://link.twitch.tv/help.}} This message has appeared frequently in this category. If a user were to follow the provided link, they would be sent to a site where they can fill out a form and submit a request for Twitch help. 

Occasionally, TPDs identified a question that was not meant for the TPD Discord server and would explain to others that they should post it on an alternative site. For example, a TPD responded, 
\begin{quote}
    I think that's outside the scope of this discord server. I think, no one can help you out here. Just try to explain this bug in detail in the ticket. The formal support is the ticket system:   https://help.twitch.tv/customer/portal/emails/new.
\end{quote} 
This responding TPD provided a link to a Twitch support ticket entry, but other TPDs have directed users to other Discord servers or websites that may offer more assistance. 

TPDs would also redirect these support seekers to other channels with relevant resources.
For example, a TPD said, \say{\textit{Which isn't going to happen. So you'd be better off requesting the data/endpoints you want through UserVoice so Twitch knows that there's a demand for it to be added to Helix.}}  In this case, the TPD suggested reporting through UserVoice to indicate their demand and raise awareness of Twitch. Another TPD said all TPDs should contact Twitch first to obtain a developer agreement to use the Twitch API. A TPD recommended, \say{\textit{Check if spotify has a support channel for developers, otherwise I would recommend to ask on stackoverflow or other support channel.}}

\subsubsection{Socializer}

TPDs occasionally used the server as a social space to share low-stakes instances, such as temporarily running out of ideas, perceived limited value in pursuing an additional task, diminished motivation in non-project activities (e.g., recreational video game play), or social gatherings. 
A TPD explained why not completing a task: \say{\textit{as I said, there's not enough motivation to go and do that because I have enough other stuff. Plus I don't have a spare GPU for passthrough.}} 

TPDs discussed social events such as Twitch meet-ups at TwitchCon, a convention that all Twitch stakeholders would like to attend semi-annually. For example, TPDs discussed their travel plans due to the meet-up cancellation, and a TPD explained, \say{\textit{There are some that are still going and doing meetups, as they have non-refundable plane tickets/hotels, it just won't be in any sort of formal Twitch capacity.}} The TPDs explained that they could still meet at the convention in an alternative way. 

\subsubsection{Conflict Facilitator}
Conflict facilitators were TPDs who attempted to diffuse conflict in the discussion when it got repetitive and heated.
Sometimes arguments broke out on the Discord server, and TPDs tried to resolve them. For instance, 
\begin{quote}
One simple question: Did you use the form at <https://www.twitch.tv/p/en-gb/security/> to report them? Yes or No? Regardless, it is unnecessary to tell other members here to "shut up" when they are contributing to a healthy discussion and trying to help you with your questions.
\end{quote}
In this example, a TPD responded to a complaint by another TPD about a different incident, providing them with a resource to take action. The TPD also attempted to keep the server a positive space by acknowledging the rude behavior of the TPD who gave the initial complaint. 

In the case of another disagreement, a TPD pointed out the lack of direction in their discussion:
\begin{quote}
It's something we are aware of but all this discussion has done is gone around in a circle. The TLDR is just deal with it. Twitch is aware of it and handling it in their own way And we dunno what Twitch is or is not doing about it or how it does or does not factor into other stuff on the platform.
\end{quote}
The TPD tried to bring the conversation to a close and explained that the issue they were having with Twitch would not be solved within the argument. Instead, Twitch would handle the issue just as they had handled other issues brought to their attention.

Other TPDs employed alternative methods to mitigate conflicts within the chat. For instance, a TPD could summarize the points of an unhelpful discussion to move past the argument and get the involved parties to make progress in their work:

\begin{quote}
    Time to calm this conversation down a little. So main points! 1) @XXX gab a dev and get them in here... we can then answer their questions 2) @XXX do we need to implement a swear filter for you? £) Yes i used the £ because I was pressing shift  4) No more relaying please 5) Some devs do work for free, on big projects.
\end{quote}
Other summaries of conflict might explain the types of conflicts commonly reported by TPDs. More specifically, \say{\textit{Today on the dev forums, we have spam, we have people posting stuff we can't help with and someone breaking every rule in the book.}} This TPD showed how spam and inappropriate posts were some issues within the server.

TPDs may provide a different type of summary, but with a purpose other than resolving an argument. For instance, a TPD summarized the lighthearted suggestions they had for expanding the TPD community: \say{\textit{so, 1) twitch needs to open a big dev office here 2) we need more get togethers with the german dev community.}} This type of summary alternatively did not indicate there was conflict within the server, and rather listed the main tasks the TPD would like to see completed in the future. Summaries in the Discord server were used so that other TPDs can reach the conclusion of the discussion without having to sort through other comments. 

\subsubsection{Complimenter}
TPDs would compliment others for a variety of reasons. They would compliment other TPDs on their creativity when creating projects or thinking of solutions: \say{\textit{omg lol, i love how creative you guys are xD, i also watch one piece btw.}} This compliment could be catered to a specific group of TPDs, coming up with creative solutions, or even an off-topic conversation.  

TPDs would thank those who provided them with advice, including identifying their mistakes, suggesting platforms for project completion, offering solutions to problems, and providing feedback on projects. TPDs appreciated the help they received on this Discord server and consistently expressed their gratitude, fostering a healthy TPD community. A TPD stated, 
\begin{quote}
    Just wanted to say this discord has been awesome while developing my first Twitch Extension. I have to say I've had so many questions answered by simply searching this TwitchDev discord, never having to post a question myself. Thanks for all who answer questions, like XXX, XXX, and XXX to name a few. Appreciate ya.
\end{quote}  
Similarly, another TPD stated, \say{\textit{again thanks @XXX and @XXX for your helpful feedback and @XXX … seem like you talking about something, I never knew about the invite only tab. so i'll look more into this.}} This TPD called out specific TPDs that had directly helped them. 
Additionally, they might announce that an issue has been resolved and thank the support provider, providing an explanation.

\section{Discussion}
This study explores Twitch TPDs' informal support practices on both the seeking and the provision sides. We identified the high-level types of support-seeking and the diverse roles of support providers. The support-seeking practices are clearly tied to the core space for which they work and should have received formal support. However, they had to venture into peripheral spaces to seek informal support. We argue that TPDs' support-seeking practices in the Discord community reflect their dependence on Twitch. These dependencies can lead to platform labor; the space migration may cause confusion and uncertainty, thereby exacerbating the platform labor. Although TPDs played diverse roles in support provision, there is a need to align roles and responsibilities across the two platforms to bridge gaps and break the problem in the loop, thereby facilitating collaboration between the TPDs and the core platform. We discussed TPDs' support seeking and provision from four perspectives: dependence on Twitch, platform labor on Twitch, migration dilemma from Twitch to Discord, and flexible roles on Discord, as summarized in \autoref{tab:my-table}.

\begin{table}[h]
\centering
\caption{An overview of the discussion points with corresponding design implications, and the stakeholders should implement these designs.}
\resizebox{\textwidth}{!}{%
\begin{tabular}{lll}
 \toprule
\textbf{TPD's Support Seeking and Provision} & \textbf{Design Implications} & \textbf{Who should mainly implement} \\  \hline
Dependence on Twitch & \begin{tabular}[c]{@{}l@{}}Design to balance the platform control \\ and TPDs’ autonomy.\end{tabular} & Twitch \\
Platform Labor on Twitch & \begin{tabular}[c]{@{}l@{}}Design to improve their visibility, \\ recognition, and reduce precarity\end{tabular} & Twitch \\
\begin{tabular}[c]{@{}l@{}}Migration Dilemma\\  Between Twitch and Discord\end{tabular} & \begin{tabular}[c]{@{}l@{}}Design to balance the boundary support \\ between the core and peripheral platform\end{tabular} & \begin{tabular}[c]{@{}l@{}}Twitch or Discord \end{tabular} \\
Flexible Roles on Discord & \begin{tabular}[c]{@{}l@{}}Design for roles and responsibilities \\ to bridge the two platforms\end{tabular} & Discord  \\  \bottomrule
\end{tabular}%
}

\label{tab:my-table}
\end{table}

\subsection{TPDs' Support Seeking in the Discord Community as Their Reflection of Dependence on Twitch}
Similar to FOSS developers, TPDs often encounter development challenges. In contrast, Twitch's formal channels often fail to provide timely responses, forcing TPDs to turn to the Discord community for support. TPDs' support-seeking practices can be a mix of social (learning and hardware), technical (account, support speed, and development questions), and political (rules and guidance) inquiries on the Discord server. These inquiries reflect their high dependency on the Twitch platform. 

The technical development work of TPDs spans multiple areas, including front-end deployment and back-end database design, extension development, and anti-bot strategies. The lack of formal support from Twitch makes it challenging for TPDs to acquire the required knowledge, as prior work has shown that the API and related documents alone are insufficient for effective development \cite{fan_why_2021}. In this situation, relying on technical inquiry has become a necessary way for TPDs to fill knowledge gaps and accumulate practical experience. Unlike FOSS developers, who are often volunteers with blended motivations \cite{jahn_blending_2025} and participatory decision-making \cite{filippova_effects_2016}, Twitch has the final say in all decisions but lacks transparency and is delayed. 

The lack of clarity in policy and legal advice directly contributes to the complexity of TPDs' work. Prior work shows that fewer than half of the platforms are complete in terms of policy composition \cite{schaffner_community_2024}, leading to TPDs' unclear understanding, especially when policy updates occur. They often consult platform policies and seek legal advice to ensure that their actions do not violate platform rules or relevant laws, which may result in issues such as copyright infringement complaints or website shutdowns. Such behaviors may restrict their creative freedom during the development process, as prior work shows \cite{freeman_understanding_2023}. 
% \cite{chakraborti_nlp4gov_2024,zhang_data_2024,schaffner__2024}.

TPDs communicate with their peers, ensuring their work continues, and sometimes, they complain together. Through sharing these common hardships, TPDs form social attachments to the community, constructing identities and navigating the community's dynamics \cite{ducheneaut_socialization_2005}, which serves as an important precondition for them to practice reciprocity \cite{kim_receivers_2022}.
This kind of socioemotional inquiry is positive for community building but is rooted in technical and regulatory inquiries.  

We argue that the TPDs' support-seeking practices reflect their dependence on Twitch. However, we cautiously argue that TPDs should reduce their dependence on formal platform support to achieve more efficient development practices, which differs from other platform workers who emphasize independence and great autonomy \cite{alvarez_de_la_vega_design_2022}. Even with this dependence, developers can strategically reshape their work to gain advantages by structuring boundary resources \cite{tian_how_2025}. Thus, it is better for Twitch to consider how to balance the platform's control and the development of TPDs, such as lowering the entry barrier for developers by providing user-friendly APIs \cite{upreti_advancing_2025}, improving platform compatibility with diverse deveopement experience \cite{tian_platform_2022}, integrating personal values with societal impact into the development \cite{nurain_hugging_2021}, and establishing transparent platform guidelines \cite{bulat_psychology_2024} to ensure that TPDs are not distracted by concerns over micromanagement, allowing them to focus solely on development itself \cite{chakraborti_nlp4gov_2024,zhang_data_2024}. These suggestions aim to expedite the support inquiries identified in this study, enhancing TPDs' autonomy while maintaining a certain level of control from Twitch.

\subsection{TPD's High Dependence on Twitch as Platform Labor}

Platform labor in this study refers to the labor that TPDs are involved in conducting mediated work for the platform ecosystem. This high dependence essentially reflects the power asymmetry between platforms and TPDs, similar to gig workers who are in a power structure dominated by the platform and are highly dependent on the platform's resource allocation and policy, often needing to build resilience to address these challenges 
\cite{jarrahi_platformic_2020,yao_together_2021, schaffner_community_2024}. This dependence poses challenges to stakeholders with less power, such as TPDs 
\cite{bruun_coordination_2025, hellman_ohhh_2025}. At key levels such as policy formulation and legal application, platforms, by virtue of their resource advantages and the power to formulate rules, hold greater discourse power and dominance \cite{gilbert_towards_2023}. In contrast, TPDs are in a relatively disadvantaged position and highly rely on platforms' infrastructure and guidance. This power structure leaves TPDs with insufficient bargaining power and decision-making autonomy when faced with ambiguity. This situation is similar to that of gig workers on centralized platforms, who face numerous obstacles and exploitation; they need additional labor (e.g., emotional labor) to manage these issues, rather than focusing on the work itself \cite{yao_peer_2024, raval_standing_2016}. 

However, the regulatory and opaque management, accompanied by frequent infrastructure updates and delayed asynchronous communication, can also cause precarity. Although gig workers faced challenges caused by the platform, such as reduced visibility \cite{zhang_aura_2025}, the TPDs, as peripheral compared to professional developers, and not even contractors with the core platform, experience even more precarity, resonating with prior work indicating that Twitch generally consolidates power over its ecosystem during its evolution, engendering precariousness in its dependents like TPDs \cite{partin_bit_2020}. Meanwhile, Twitch provides limited channels to collect their voice, such as the Twitch Voice forum and email list, but lacks a specific design to better distribute its resources to the TPDs. TPDs need to spend more time and incur higher costs to manage this additional labor (e.g., searching and waiting) than gig workers on other platforms. These voluntary efforts occurred in a peripheral space that might not even be aware of by the core platform, exacerbating invisibility and precarity. In this sense, we extended prior work by encouraging research on community-building for gig workers who lack policy and social support, as well as fear of damaging their relationship with the platform \cite{hsieh_co-designing_2023}, from the peripheral platform's informal support and labor perspective.  

Platforms must carefully manage both internal and external value creation \cite{parker_platform_2017}. Online systems designed for social interactions should connect diverse activities \cite{tixier_counting_2016}. We argue that Twitch TPDs play a vital role in the core space (Twitch) and that their platform labor on Discord should be acknowledged through effective design. Twitch should highlight TPDs' invisible labor on the core platform, ensure their contributions are recognized, and reduce precarity and frustration while fostering collaboration between TPDs and Twitch.

\subsection{TPDs' Support Seeking and Provision Dilemma}

Platform workers, such as gig workers and content creators, experience platform labor, which involves managing algorithms \cite{mohlmann_algorithmic_2021} that continually cause dishonor and uncertainty \cite{grohmann_platform_2022, stelmaszak_when_2025}, and complain that platforms show limited care for their individual labor on managing their activities on the platform \cite{ma_i_2022}. They might bypass the platform's algorithmic management by forming other online communities for support and task-specific collaboration \cite{saether_workers_2023}. In this sense, TPDs are similar to other platform workers, who migrate from the core platform to the peripheral space for support.

The push-pull migration model is used to explain human migration \cite{lee_theory_1966} and has been applied in online communities (e.g., \cite{zhang_understanding_2009, hsieh_post-adoption_2012}). When users experience negative aspects in one online community, such as low information quality, support failure, and poor service performance, which serve as push effects for migration, they may migrate to another online community with high information quality, socialization, and enjoyment, which serve as pull effects \cite{hou_understanding_2019}.

The lack of formal support from the Twitch platform is typical of push effects, characterized by slow responses, ambiguous information, and instructions. The Twitch formal support channels (e.g., UserVice, email) function more like passive data repositories, waiting for TPDs to actively search for information, and fail to provide effective assistance. Meanwhile, there is a lack of synchronized and up-to-date information from the platform, along with an inadequate coordination mechanism to meet TPDs' needs, driving them to other spaces. 

The Discord community serves as a gathering space for informal support among TPDs with diverse backgrounds, offering various levels of social and technical questions with corresponding support roles and answers, as well as socializing and civil discussions with appreciations and satisfactions. Prior work shows that developers would like to contribute if their values align with the community value
\cite{maruping_developer_2019}, and use communication channels to maintain group awareness \cite{gutwin_group_2004}. In this sense, the Discord community, as an alternative, presents a list of pull effects that attract TPDs to join and sustain the online community.   

The migration from the core to the peripheral space is a bidirectional process. Most TPDs must switch between these two platforms to find the best answers. When they cannot get information from formal spaces, they turn to informal spaces to seek support. However, the Discord community may redirect them to the core space where they initially started, creating a circular feedback loop (Twitch-Discord-Twitch). This closed loop results in considerable time spent transferring problems between spaces, leading to confusion and uncertainty. Such uncertainty can impact work efficiency and undermine the value of knowledge sharing within the entire community \cite{jahn_blending_2025}.  

We argue that both core and peripheral spaces should complement each other, rather than creating tension between adhering to one or the other, or forcing TPDs to be pushed back and forth. Prior work shows that formal and informal support among developers helps close knowledge gaps \cite{hwang_adopting_2024}. Thus, there needs to be a push-pull effect balance between these spaces, with clear boundary support to mitigate uncertainty and the knowledge gap. For example, the core space (Twitch) can maintain high information quality and rapid responses with infrastructure or system updates. The peripheral space (Discord) can facilitate socializing to welcome newcomers \cite{singh_types_2011}, allow for the sharing of experiences (e.g., emotional venting) \cite{progga_just_2023, progga_understanding_2023}, and maintain civil discussions among stakeholders through the exchange of ideas and innovation \cite{cai_third-party_2024}.

\subsection{TPDs' Support Provision with  Flexible Roles In Discord Community }

Support providers played diverse roles within the Discord community, ranging from instructor, explainer, adviser, support redirectioner, socializer, conflict facilitator, and complimenter, among others.
The diverse roles reflect the flourishing of the informal space, which relies on community effort and aligns broadly with Kraut et al.'s seminal work on design principles for building a successful online community \cite{kraut_building_2016}. These principles range from promoting positivity (e.g., attracting and sustaining newcomers, encouraging commitment and contribution) to curbing negativity (e.g, facilitating regulation and civil discussion). 

For example, TPDs take on roles such as instructors, advisors, and explainers to assist community members. They focus on sharing constructive feedback and valuable information, often based on their personal experiences, to help others overcome challenges on platforms like Twitch, echoing prior work showing community members can serve as recommenders and troubleshooters to sustain community building \cite{hwang_adopting_2024}. The support redirectors are those who, despite lacking the resources to provide direct assistance, recommend alternative sources of help, demonstrating their commitment to the community. Additionally, roles like socializers and complimenters are essential for attracting newcomers and fostering connections within the group, echoing prior work showing that members continue to participate in online communities due to the obligations of reciprocity and the ties they have formed with other community members \cite{yao_join_2021}. The conflict facilitators contribute to the regulation and harmony within the community, similar to OSS project management \cite{ferreira_shut_2021}. 
The frustrations and challenges caused by dependence on Twitch were recast into positive roles that TPDs played, similar to prior work showing that collaboration to share labor with support can recast a negative experience into a positive one \cite{hwang_adopting_2024}. Additionally, the Discord community offers a level of privacy that other public social media platforms, such as X, lack, as prior work has shown that private channels can promote self-disclosure to facilitate socialization and enhance social support \cite{hoefer_bridging_2022}.

Although these roles reflect the flourishing of the Discord community, with a focus on support provision, the seeking side highly depends on the core platform Twitch. These informal support provisions can't fundamentally solve the problems caused by the core space. Thus, there must be some other roles to bridge the gap between the core and peripheral spaces. Prior work shows that some developer communities have publicizers who actively promote and share new features to the public, ensuring that tool users are informed about updates  \cite{haraty_online_2017}. The core members who actively engage in the support provision are the core users that a community design should focus on to sustain the online community \cite{introne_sociotechnical_2016}. Similarly, the Discord community could benefit from having a dedicated role that aggregates problems and effectively communicates with Twitch's officials. Although Twitch staff members participate in the community, they are not encouraged to engage in the support-seeking and provision process. However, there is potential for these staff members to act as mediators, sharing and addressing support-seeking issues, or for other TPDs to take these responsibilities. 

\subsection{Limitations and Future Directions}

There are several limitations for future work. First, this study has focused only on Twitch TPDs in the Discord community. TPDs on platforms such as Figma, Google, and iOS applications may also present similar problems \cite{vaccaro_at_2020}. For example, there are changes to APIs (new versions do not support the functions of older versions), or the development tools or APIs are owned by multiple companies \cite{cai_third-party_2024}. Future research can compare the problems TPDs face across environments to develop a more comprehensive theoretical framework. Second, some TPDs have reported that certain policies restrict their usage, highlighting the contradiction between the availability of technical resources and the constraints imposed by platform policies \cite{fiesler_understanding_2015}. Future research can explore the power balance between platform owners and TPDs. Third, because the data focus on the ``Lobby'' channel and do not consider other specific channels, the analysis primarily uses qualitative categories. Future work can combine all channels with a large data corpus and quantitative methods to explore the dynamics and evolution of behaviors and roles \cite{im_deliberation_2018, arazy_how_2017}.

\section{Conclusion}

In this study, we employed a mixed-methods approach to investigate Twitch TPDs' support practices within a Discord community. We show that informal support is deeply shaped by TPDs' dependence on Twitch, turning support-seeking and coordination across spaces into a form of platform labor. When formal support is limited, TPDs often migrate to peripheral community spaces to maintain development progress, which can intensify uncertainty and increase additional labor. While flexible community roles sustain support provision, our findings suggest the need for clearer bridges between formal and informal spaces to reduce unnecessary labor and improve collaboration in platform ecosystems.

\begin{acks}
Thanks to the undergraduates, Fiona Rubino, Sydney Graver, and Oluwaseni Johnson, for supporting the data coding process.  
\end{acks}

\bibliographystyle{ACM-Reference-Format}
\bibliography{TPD, sample-base}
% \appendix
\end{document}